\documentclass[english,prd,twocolumn,showpacs]{revtex4}
\usepackage[latin1]{inputenc}
\usepackage{fontenc}
\usepackage{babel}
\usepackage{graphicx,subfigure}
\usepackage{amsmath,amssymb}
\newcommand{\hb}{\mathcal{H}}

\begin{document}

\title{Entropy of gravitationally collapsing matter in FRW universe
models}
\date{\today}
\author{Morad Amarzguioui}
\email{morad@astro.uio.no}
\affiliation{Institute
    for Theoretical Astrophysics, University of Oslo, PO Box 1029
    Blindern, 0315 Oslo, Norway}

\author{Øyvind Grøn}
\email{oyvind.gron@iu.hio.no}
\affiliation{Oslo College, Faculty of Engineering, Cort Adelersgt.
30, 0254 Oslo, Norway}
\affiliation{Department of Physics, University of Oslo, PO Box 1048
    Blindern, 0316 Oslo, Norway}

\begin{abstract}
We look at a gas of dust and investigate how its entropy evolves with
time under a spherically symmetric gravitational collapse. We treat the
problem perturbatively and find that the classical thermodynamic
entropy does actually increase to first order when one allows for
gravitational potential energy to be transferred to thermal energy
during the collapse. Thus, in this situation there is no need to
resort to the introduction of an intrinsic \emph{gravitational
entropy} in order to satisfy the second law of thermodynamics.
\end{abstract}

\pacs{95.30.Sf, 95.30.Tg}

\maketitle

\section{Introduction}
Today there is broad consensus among cosmologists that the
configuration of energy in the early universe was very homogeneous and
isotropic. Observations of the temperature variations in the cosmic
microwave radiation have also shown that the Universe was in a state
close to thermodynamic equilibrium $400\,000$ years after Big Bang,
with relative temperature and density variations of the order
$10^{-5}$ \cite{spergel03}.

Naively, one expects the entropy in a gas to be higher the more
homogeneously distributed its density and temperature is. Thus, the
early universe described above should be one of near maximal entropy,
since it differs only by a small fraction from one of total
homogeneity in density and temperature. However, due to gravity, small
inhomogeneities start to grow and eventually end up forming structures
such as galaxies, stars, planets, planetary clouds etc. This evolution
is in the direction of greater inhomogeneities both in energy density
and temperature, which according to the argument above, appears to
violate the second law of thermodynamics by decreasing the entropy.
Obviously, something must be wrong with this picture, since we
consider the second law of thermodynamics to be a basic law of physics
and it should therefore not be violated. 

A possible solution to this apparent paradox comes from considering
the quantity known as gravitational entropy. This was introduced by R.
Penrose in the 1977 in connection with his study of the properties
of the initial singularity of the universe
\cite{penrose77,penrose79,penrose81}. It is a quantity which can be
interpreted as an entropy intrinsic to the gravitational field. It
takes into account the attractive nature of gravity and increases as a
gas collapses under the influence of gravity. This allows one to
define a general entropy which is the sum of the ordinary
thermodynamic entropy and this new gravitational entropy. If the sum
of the two types of entropy increases during gravitational collapse of
a gas, the second law of thermodynamics will then be preserved.


In this paper we will show that the thermodynamics entropy of a
collapsing gas does actually increase, which allows us to explain the
collapse without introducing the gravitational entropy. We look at a
perturbed ideal gas in a FRW background and consider changes in its
classical entropy up to first order in the energy density. We find
that the increase in the thermal energy which comes from  potential
energy released in the collapse actually makes the total thermodynamic
entropy increase, even though the temperature inhomogeneity increases.

The structure of this paper is as follows. In section 2 we look at a
simplified model consisting of ideal particles in a box and explain
why one would expect the thermodynamic entropy to decrease as the
inhomogeneities increase. In section 3 we introduce a tool which we
will need when considering the growth of small inhomogeneities, namely
cosmological perturbation theory. In section 4 we derive an
expression for the thermodynamic entropy of a gas in an expanding
universe. In section 5 we specialize to spherically symmetric
collapsing gases and arrive at our main result. Finally, section 6
contains a summary and our conclusion.

\section{Simple picture: Ideal gas in a box} \label{sec:box}
In this section we look at gas confined to a box and show that its
entropy is maximal when the density is homogeneous and the temperature
is the same everywhere.

\begin{figure}
\begin{center}
\includegraphics[width=7.2cm]{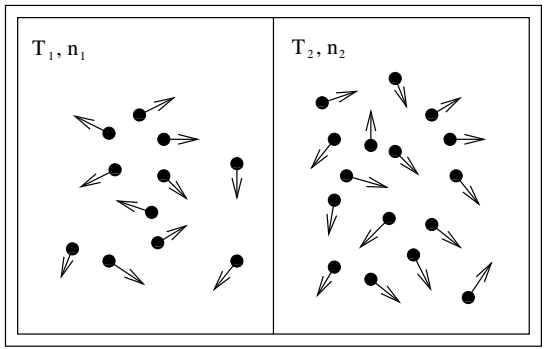}
\end{center}
\caption{\label{box}An isolated box consisting of two separated chambers with
particles of different temperatures and densities.}
\end{figure}

Consider an isolated box that is divided into two chambers of equal
volume. Each of these contains a gas of the same type of particles
with different temperatures and densities, as illustrated in
Fig.~\ref{box}. We will look at two different scenarios: 1) when the
temperature in the chambers is the same but the density is different.
And 2) when temperature is different, while the density is the same.
We then compare these to a third scenario, in which we remove the wall
between the two chambers so that we have only one gas with just one
temperature and one density.

First, we need the expression for the entropy of an ideal gas
consisting of $N$ ideal particles in a volume $V$ \cite{kittel80},
\begin{equation} \label{SIdeal}
	S_{\mbox{\tiny{ideal}}}=Nk_B\ln{\left[\frac{V}{N}
  \left(\frac{mk_BT}{2\pi\hbar^2}\right)^{3/2}e^{5/2}\right]}\,.
\end{equation}

Let us look at the third scenario first. Let $T$ be the temperature in
the gas and $N$ the total amount of particles in the box. The entropy
in the box in this scenario, which represents the totally homogeneous
case, is then
\begin{equation} \label{S3}
	S_3=Nk_B\ln{\left[\frac{KT^{3/2}V}{N}\right]}\,,
\end{equation}
where we have defined the constant
\begin{equation} \label{DoK}
	K=\left[\frac{mk_BTe^{5/3}}{2\pi\hbar^2}\right]^{3/2}.
\end{equation} 

How does the entropy differ from this in the other two, inhomogeneous
scenarios? In the first scenario the temperatures are the same, but
the number of particles in the two chambers is different. The total
number of particles is conserved, so we have that $N_1+N_2=N$, where
$N_1$ and $N_2$ are the numbers of particles in the two chambers
respectively. Furthermore, we assume that the two chambers have the
same volume $V/2$. The total entropy in the box is then the sum of the
entropies in the two chambers:
\begin{align}
	S_1&=N_1k_B\ln{\left[\frac{KT^{3/2}V}{2N_1}\right]}+
	N_2k_B\ln{\left[\frac{KT^{3/2}V}{2N_2}\right]}\notag\\
	&=Nk_B\ln{\frac{KT^{3/2}V}{2}}-\left(N_1k_B\ln{N_1}
	+N_2k_B\ln{N_2}\right)\label{S1}
\end{align}
A transition from the totally homogeneous scenario to this results in
an entropy difference which is
\begin{equation} \label{dS3}
	\Delta S = S_1-S_3=-Nk_B\bigg(\ln{2}+x\,\ln{x}+(1-x)\ln{(1-x)}
	\bigg)\,,
\end{equation}
where we have defined $x\equiv N_1/N$. We can restrict $x$ to the
interval $0<x<1/2$ without any loss of generality. It is then a simple
task to verify that $\Delta S<0$. Thus, based on this simple picture,
we can conclude that an increase in inhomogeneity of the density of an
ideal gas leads to a reduction of the entropy.

Now, let us look at the second scenario, where there is a temperature
difference between the two chambers which both contain the same amount
of particles, $N/2$. The temperatures of the two chambers are $T_1$
and $T_2$ respectively. The total entropy of the box in this scenario
is 
\begin{equation} \label{S2}
	S_2=Nk_B\left(\ln{\frac{KV}{N}}+\frac{3}{4}\ln{(T_1T_2)}
	\right)\,.
\end{equation}
If we imagine the gas first being in the totally homogeneous state of
scenario 3 and then changing into the thermally inhomogeneous state of
scenario 2, the entropy difference will be
\begin{equation} \label{dS2}
	\Delta S = \frac{3Nk_B}{4}\ln{\frac{T_1T_2}{T^2}}\,.
\end{equation}
Assume now that the thermal energy is conserved in this transition,
i.e. that the average temperatures in the two scenarios are the same.
This means that $T_1+T_2=2T$. Under this assumption, when does this
entropy difference become non-negative? We see that this will be the
case when
\begin{equation}
	\frac{4T_1T_2}{(T_1+T_2)^2}\geq1 \quad\mbox{ i.e. }\quad
	(T_1-T_2)^2\leq0\,.
\end{equation}
This is only satisfied when $T_1=T_2=T$, in which case there will be
an equality between the left and the right hand side. If $T_1$ and
$T_2$ are different, i.e. there is a temperature difference between
the two chambers, the entropy difference in \eqref{dS2} will be
negative. Thus, an increase in temperature inhomogeneity will result
in a decrease in the entropy. 

However, we must not forget that we have assumed that the thermal
energy is conserved, just as we assumed that the particle number is
conserved. For this simple example of particles in an isolated box
both these assumptions will be true. For a gravitationally collapsing
gas, however, this need not be true. The total mass, which is the
equivalent of the total particle number in the box example, must
obviously be conserved so we should expect the entropy to decrease as
the energy density becomes more inhomogeneous. But the thermal energy
will not be conserved as the gas undergoes a gravitational collapse.
As the energy density near the overdensity increases, the potential
energy of the inward falling portions of the gas is converted to
thermal energy. In other words, there will be an increase in the
thermal energy of the gas, which tends to increase the entropy. 

To summarize, the entropy change in a gravitationally collapsing gas
can be ascribed to two different effects, namely a decrease due to
increasing density and temperature inhomogeneities and an increase due
to increasing temperature. As we will show, using first order
perturbation theory, the sum of these two effects yields an increasing
total entropy when one assumes that all the loss in potential energy
is converted into thermal energy.

We start by reviewing the basics of scalar perturbation theory.

\section{Scalar perturbation theory}
We will only consider scalar perturbations since these are the only
ones that give rise to gravitational collapse. For a more detailed
review of perturbation theory, the interested reader is referred to
the standard references \cite{mukhanov92,ma95,brandenberger03}. We
assume that the universe is occupied by matter in the form of a
perfect fluid with no anisotropic stress. The energy density
inhomogeneity is described as a linear perturbation to a flat, matter
dominated FRW universe. This allows us to write the perturbed metric
in terms of only one perturbing function $\Phi$, the so-called Bardeen
potential: 
\begin{equation} \label{lineelement}
  ds^2=a^2(\eta)\left\{(1+2\Phi)d\eta^2-(1-2\Phi)\delta_{ij} dx^idx^j
  \right\}\,,
\end{equation}
where $\eta$ is conformal time and $a$ is the scale factor of the
universe.

The energy-momentum tensor for the matter content is written as a
homogeneous zeroth order term plus a non-homogeneous first order
perturbation:
\begin{equation} \label{unpertmattert}
  T^{\mu}_{\nu}=^{(0)}\!\!T^{\mu}_{\nu}+\delta T^{\mu}_{\nu}\,.
\end{equation}
Using the definition for the energy-momentum tensor of a perfect
fluid, the equation of state for matter, and the four-velocity
identity $u^{\mu}u_{\mu}=1$, we can write the components as:
\begin{align}
  ^{(0)}\!T^{0}_{0}&=\rho_0&\delta T^{0}_{0}&=\delta\rho\label{t00}\\
  ^{(0)}\!T^{0}_{i}&=0&\delta T^{0}_{i}&=-\rho_0a\delta u^i\label{t0i}\\
  ^{(0)}\!T^{i}_{j}&=0&\delta T^{i}_{j}&=0\label{tij}
\end{align}
where $\rho_0$ is the average density of the fluid, $\delta\rho$ is
the density perturbation and $\delta u^i$ the velocity perturbation.
In the expressions above and throughout this paper we use the
convention that Latin indices run over spatial components only, while
Greek indices run over both space and time components.

We require the Einstein equation for the fluid to be satisfied
independently for each order in the perturbation. This gives us the
following zeroth order equations
\begin{equation} \label{einstein00}
  \hb^2=\frac{8}{3}\pi Ga^2\rho_0
\end{equation}
and
\begin{equation} \label{einsteinij}
  \hb^2+2\dot{\hb}=0\,,
\end{equation}
where $\hb=\frac{1}{a}\frac{da}{d\eta}$ and the dot denotes
differentiation with respect to $\eta$. The first order equations are
\begin{align}
  &\nabla^2\Phi-3\hb(\dot{\Phi}+\hb\Phi)=\frac{3}{2}\hb^2\delta
  \label{pert00}\\
  &\left\{\dot{\Phi}+\hb\Phi\right\}_{,i}
  =-\frac{3}{2}\hb^2a\delta
  u^i\label{pert0i}\\
  &\ddot{\Phi}+3\hb\dot{\Phi}+(\hb^2+2\dot{\hb})\Phi=0\label{pertij}
\end{align}
where we have defined the density contrast as
$\delta\equiv\delta\rho/\rho_0$, and
$A_{,j}\equiv\frac{\partial A}{\partial x^j}$.
The zeroth order equations are the ordinary FRW equations for a matter
dominated, flat universe expressed in conformal time instead of the
usual comoving time. The solution to these equations is 
\begin{equation}
  a=\left(\frac{\eta}{\eta_0}\right)^2\,\quad \mbox{and}\,\quad
  \rho_0=\rho_{00}\left(\frac{\eta_0}{\eta}\right)^6\,,
\end{equation}
where we have defined $\eta_0$ such that $a(\eta_0)=1$, and
$\rho_{00}$ is the energy density at $\eta=\eta_0$, which can be
written as
\begin{equation} \label{defOfRho00}
	\rho_{00}=\frac{3}{2\pi G\eta_0^2}\,.
\end{equation}
We shall later need the relation between conformal and comoving time.
Using the definition $a(\eta)d\eta=dt$, we can write this as
\begin{equation} \label{reletat}
  \eta=\eta_0\left(\frac{t}{t_0}\right)^{\frac{1}{3}}\,,
\end{equation}
where $t_0$ is the initial comoving time that corresponds to $\eta_0$.
These are related via the expression
\begin{equation} \label{relInitTimes}
	t_0=\frac{\eta_0}{3}\,.
\end{equation}
Let us introduce a new dimensionless time parameter $\tau$, which
measures time relative to the initial time $t_0$, i.e.
\begin{equation} \label{taudef}
  \tau\equiv\frac{t-t_0}{t_0}\,.
\end{equation}
Using this new time parameter we can write the scale factor and the
unperturbed energy density as
\begin{equation} \label{arhotau}
  a(\tau)=(1+\tau)^{\frac{2}{3}}\,\quad \mbox{and}\,\quad
  \rho_0(\tau)=\rho_{00}(1+\tau)^{-2}\,.
\end{equation}

Next, we solve the first order equations. We start with equation
\eqref{pertij}, which doesn't couple to the other two equations,
and obtain the metric perturbation $\Phi$. The remaining perturbing
functions $\delta$ and $\delta u^i$ are then obtained by a simple
substitution of $\Phi$ into equations \eqref{pert00} and
\eqref{pert0i}. Disregarding solutions that decrease with time, we
can write the perturbations as:
\begin{align}
  \Phi(\mathbf{x},\tau)&=f(\mathbf{x})\label{phi}\\
  \delta(\mathbf{x},\tau)&=\frac{1}{6}\eta_0^2(1+\tau)^{2/3}\,
  \nabla^2f(\mathbf{x}) -2f(\mathbf{x})\label{delta}\\
  \delta u^i(\mathbf{x},\tau)&=-\frac{\eta_0}{3}\frac{d
  f(\mathbf{x})}{d x^i}(1+\tau)^{-1/3}\label{deltaui}
\end{align}
where $f(\mathbf{x})$ is an arbitrary function of spatial coordinates
only. In order for these perturbations to be physically acceptable
they must vanish at infinity, i.e.
\begin{equation} \label{pertinf}
	\lim_{\|\mathbf{x}\|\to\infty}f(\mathbf{x}) =
	\lim_{\|\mathbf{x}\|\to\infty}\delta(\mathbf{x},\tau) =
	\lim_{\|\mathbf{x}\|\to\infty}\delta u^i(\mathbf{x},\tau)
	= 0\,.
\end{equation}
Furthermore, we must also require that the total energy density at
$\tau=0$ in the perturbed and the unperturbed universe remains the
same, which is the same as saying that the total energy must be
conserved when the perturbation is introduced. For this requirement to
be satisfied, the initial density perturbation must satisfy the
integral condition
\begin{equation} \label{InitEnergyCons}
	\int_V\!\delta(\mathbf{x},0)\,dV=0\,.
\end{equation}
Substituting for the left hand side from Eq. \eqref{delta} and using
the boundary conditions in Eq. \eqref{pertinf}, we find that the
volume integral of the metric perturbation must also vanish,
\begin{equation} \label{fcons}
	\int_V\!f(\mathbf{x})\,dV=\int_V\!\nabla^2f(\mathbf{x})\,dV
	=0\,. 
\end{equation}
This implies that the volume integral of the density perturbation must
vanish for all values of $\tau$,
\begin{equation} \label{TotEnergyCons}
	\int_V\!\delta(\mathbf{x},\tau)\,dV=0\,.
\end{equation}
What this equation says is simply that the total energy in the
perturbed universe must be the same as that in the unperturbed
universe at all times. This is nothing but a statement of energy
conservation.

\section{Entropy of a perturbed ideal gas in a FRW universe}
Equation \eqref{SIdeal} gives us the entropy of an ideal gas
consisting of $N$ distinct particles. The collapsing gas we wish to
examine, whose time evolution is given by Eq. \eqref{delta}, consists
of a continuous fluid. We must therefore rewrite the expression in
\eqref{SIdeal} into a form that we can use for a continuous fluid.
In order to do that we consider an ideal gas contained within a small
volume element $dV$. The number of particles inside this volume is
\begin{equation}
  dN=\frac{\rho dV}{m}\,,
\end{equation}
where $m$ is the mass of the particles which the gas consists of.
Inserting this expression for the particle number into Eq.
\eqref{SIdeal}, we can write the entropy associated with the volume
element in terms of the density of the fluid:
\begin{equation} \label{dS}
  dS=k_B\frac{\rho}{m}\ln{\left(\frac{mKT^{3/2}}{\rho}\right)}
  dV\equiv\sigma dV\,, 
\end{equation}
where the constant $K$ is defined in \eqref{DoK} and $\sigma$ can be
interpreted as the entropy density of the ideal, continuous gas
distribution.

We substitute the energy density of the perturbed pressureless gas for
the density which appears in this expression. The former can be
written as $\rho=\rho_0(1+\delta)$, where $\rho_0$ and $\delta$ are
given by \eqref{arhotau} and \eqref{delta} respectively. This allows
us to write the entropy density as
\begin{equation} \label{sigmaTpert}
  \sigma_T=k_B\frac{\rho_0}{m}\left\{\ln\frac{mKT^{3/2}}{\rho_0}+\delta
  \left(\ln{\frac{mKT^{3/2}}{\rho_0}}-1\right)\right\}\,.
\end{equation}
Time dependence enters into this expression via the unperturbed energy
density $\rho_0$, the density contrast $\delta$ and the temperature
$T$. For a totally homogeneous universe which contains only matter,
the temperature can be shown \cite{davies74} to scale like
$\bar{T}\sim a^{-2}$, where the bar denotes that the temperature is
that of a non-perturbed gas. In terms of the dimensionless time
parameter $\tau$, we can write the time dependence of the homogeneous
temperature as
\begin{equation} \label{TtauHom}
	\bar{T}= T_0(1+\tau)^{-4/3}\,,
\end{equation}
where $T_0$ is the temperature of the gas at the initial time $\tau=0$.

In a perturbed gas we expect there to be an additional,
non-homogeneous contribution to this temperature. Thus, we can write
the total temperature as
\begin{equation} \label{totTemp}
	T=\bar{T}(1+\Delta_T) \,,
\end{equation}
where $\bar{T}\Delta_T$ is the non-homogeneous addition to the
homogeneous temperature $\bar{T}$ resulting from the first order
density perturbation $\delta$. As we will see later, $\Delta_T$ will
turn out be too large for us to treat it as a first order
perturbation. 

The time evolution of $\Delta_T$ depends on how much
energy we assume is transferred from potential into thermal energy due
to the gravitational collapse and how this is transferred. The two
extremes are: 1) No energy is transferred and 2) All the potential
energy is transferred adiabatically into thermal energy. The most
realistic scenario would probably be somewhere in between these two
extremes, but for simplicity we will assume the latter scenario when
we calculate $\Delta_T$ explicitly in the next section. Using Eqs.
\eqref{arhotau}, \eqref{TtauHom} and \eqref{totTemp} we can write the
entropy density of the ideal, perturbed gas as
\begin{align}
  	\sigma=\frac{k_B}{m}\frac{\rho_{00}}{(1+\tau)^2}
  	\left\{\ln{\frac{mKT_0^{3/2}}{\rho_{00}}}+\frac{3}{2}
  	\ln{(1+\Delta_T)}\right.\notag\\
  	\left.+\delta\left(\ln{\frac{mKT_0^{3/2}}{\rho_{00}}}-1
  	\right)\right\}.\label{entdens}
\end{align}
The total entropy of the gas is the volume integral of this expression
over the whole of space. The volume element which appears in the
integral is given by the determinant of the spatial metric $h_{ij}$.
In Cartesian coordinates, this can be written as
\begin{equation} \label{dV}
	dV=\sqrt{|\det{h_{ij}}|}\,d^3x=(1+\tau)^2(1-3\Phi)d^3x\,,
\end{equation}
where $d^3x$ is the Euclidean volume element. This allows us to write
the entropy element as
\begin{align} 
	dS=&\sigma dV\notag\\
	=&\frac{k_B\rho_{00}}{m}
  	\left[(1-3\Phi)\left(\ln{\frac{mKT_0^{3/2}}{\rho_{00}}}
  	+\frac{3}{2}\ln{(1+\Delta_T)}\right)\right.\notag\\
	&\qquad+\left.\delta\left(
  	\ln{\frac{mKT_0^{3/2}}{\rho_{00}}}-1\right)\right]
	d^3x \label{dStot}
\end{align}
As a consistency check we can calculate the entropy inside a comoving
volume $V$ of the unperturbed FRW model, which we know to be a
constant \cite{misner73}. Using our definition of the entropy of an
ideal, cosmological gas \eqref{dStot}, we find that
\begin{align} 
	S^{\mbox{\tiny{unpert}}}=&\int_V\!\frac{k_B\rho_{00}}{m}
  	\ln{\frac{mKT_0^{3/2}}{\rho_{00}}}\,d^3x\notag\\
	=&\,\frac{k_B\rho_{00}V}{m}\ln{\frac{mKT_0^{3/2}}{\rho_{00}}}
	\label{Sunpert}\,,
\end{align}
which is indeed a constant. What about the perturbed entropy? If the
gravitational collapse is not to conflict with the second law of
thermodynamics this should increase with time, or at least not
decrease with time. The change in entropy resulting from the density
perturbation is
\begin{align} 
	\Delta S = \frac{k_B\rho_{00}}{m}\int_V\!\left[
  	\delta(\ln{\frac{mKT_0^{3/2}}{\rho_{00}}}-1)-
  	3\Phi\ln{\frac{mKT_0^{3/2}}{\rho_{00}}}\right.\notag\\
	\left.+(1-3\Phi)\frac{3}{2}\ln{(1+\Delta_T)}\right]d^3x
	\label{SpertGeneral}\,.
\end{align}
Due to the energy conservation equations \eqref{fcons} and
\eqref{TotEnergyCons}, the first two terms in this integral will
vanish. This leaves us with a contribution from the temperature
inhomogeneity only
\begin{align}
	\Delta S =& \frac{3}{2}\frac{k_B\rho_{00}}{m}\int_V\!
  	(1-3\Phi)\ln{(1+\Delta_T)}\,d^3x\notag\\
	\approx&\,\frac{3}{2}
  	\frac{k_B\rho_{00}}{m}\int_V\!\ln{(1+\Delta_T)}\,d^3x\,.
	\label{Spert}
\end{align}
In the simplified box example of section \ref{sec:box} we found that
density inhomogeneities tend to reduce the total entropy. It might
therefore seem odd that the density inhomogeneities in the
gravitationally collapsing gas don't contribute to the entropy.
The reason that the entropy decreased in the box example is that the
expression for the entropy is non-linear in the particle number (or
alternatively in the density). This leads to the result that the sum
of the entropies of two gases with different particle numbers is
generally different from that of a gas whose particle number is equal
to the sum of the particles in the two first gases. But if we treat
the problem perturbatively to first order, we force all physical
quantities to be linear in the perturbed quantities. Since the density
is conserved, as we can see in Eq. \eqref{TotEnergyCons}, the
contribution to the entropy from the density inhomogeneity must vanish
to first order.

In the next section we examine how the entropy changes when energy
is converted adiabatically from gravitational potential energy into
thermal energy for a spherically symmetric collapsing gas.

\section{Spherically symmetric perturbations}
In this section we will specialize to spherically symmetric
perturbations. Furthermore we assume that the potential energy in the
gravitational field is transferred into thermal energy adiabatically
during the collapse. The density inhomogeneity will then give rise to
a temperature gradient. If we assume that the gas is in hydrostatic
equilibrium during the whole collapse and that the gas collapses
towards the origin of the coordinate system, we can write the
temperature gradient as \cite{carroll96}
\begin{equation} \label{defofTgrad}
	\frac{\partial\Delta_T}{\partial r}=-
	C\frac{4\pi Gm}{k_B}\frac{\rho a^2}{\bar{T}r^2}
	\int_0^r \!dr'r'^2\delta(r',\tau)\,,
\end{equation}
where $C=2/5$ for an ideal monoatomic gas. To proceed further we need
to know how the density perturbation $\delta$ behaves. We must
therefore solve the differential equations \eqref{phi} and
\eqref{delta}. In order to do this we must specify initial and
boundary conditions for both the metric and the density perturbation.
The boundary conditions are stated in Eq. \eqref{pertinf}. Once we
give an initial profile for the density perturbation we can solve Eq.
\eqref{delta} at $\tau=0$ for the metric perturbation. The time
evolution of $\delta$ is then determined by reinserting the solution
for $\Phi$ into \eqref{delta} for arbitrary $\tau$.

We restrict ourselves to a general type of initial conditions where
there is only one initial overdensity, namely centered at the origin.
These profiles must satisfy the energy conservation condition in
\eqref{InitEnergyCons}. A simple density profile contained within this
class of initial conditions is 
\begin{equation} \label{initd}
	\delta(r,0)=d_0\left[1+\frac{r}{L}-\frac{1}{3}\left(
	\frac{r}{L}\right)^2 \right]e^{-\frac{r}{L}}\,,
\end{equation}
where $L$ and $d_0$ are measures of the size and the amplitude,
respectively, of the initial overdensity. The reason that we have
chosen this explicit expression for the initial density perturbation
is that it allows us to solve the differential equation analytically.
However, as we will show in the appendix, the results we obtain apply
qualitatively for all initial density profiles in the class we defined
above.

\begin{figure}
\begin{center}
\subfigure[\label{fa}]{\includegraphics[width=7.2cm]{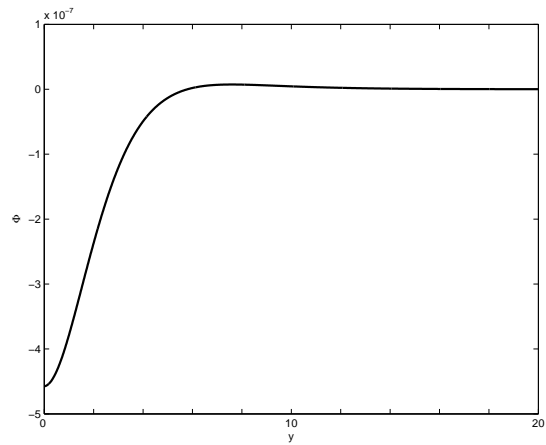}}
\subfigure[\label{fb}]{\includegraphics[width=7.2cm]{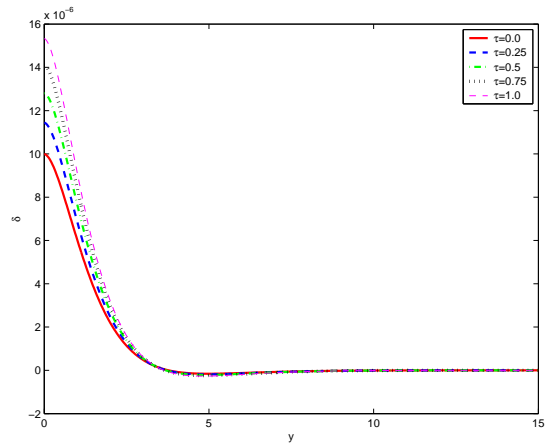}}
\end{center}
\caption{\label{fig:fd}(a) The left panel shows a  plot of the metric perturbation
that results from the initial density inhomogeneity described in Eq.
\eqref{initd}. (b) The right panel illustrates the time evolution of
the corresponding density perturbation. The curves represent the
spatial configuration of the density contrast at, from bottom to top,
$\tau=0.0,\,0.25,\,0.5,\,0.75,\,1.0$}
\end{figure}

In analogy with the time parameter $\tau$, we introduce a new
dimensionless radial coordinate $y$, which measures comoving radial
distance relative to the length scale $L$,
\begin{equation} \label{defofy}
	y\equiv\frac{r}{L}\,.
\end{equation}
The differential equation which determines the metric perturbation can
now be written as
\begin{align}
	\frac{1}{6}\left(\frac{\eta_0}{L}\right)^2\left(
	\frac{d^2}{dy^2}+\frac{2}{y}\frac{d}{dy}\right)f(y)-2f(y)\notag\\
 	=d_0\left[1+y-\frac{y^2}{3}\right]e^{-y}\,.\label{diffeqf}
\end{align}
This can be solved analytically by e.g. using the computer program
Maple. However, the analytical solution is too long for us to list
up here. Instead we illustrate the solution by plotting it in
Fig.~\ref{fa}. The amplitude of the initial perturbation was chosen to
be $d_0=10^{-5}$. This corresponds to the amplitude of the density
perturbations in our own universe at the time of recombination,
$t_0=400\,000$, which we choose as the initial time of our
perturbation. 

The time evolution of the density perturbation depends on the size of
the perturbation. If $L$ is sufficiently large, the ratio
$\frac{\eta_0}{L}$ will be so small that $\delta(y,\tau)$ remains
essentially constant in time, which can seen directly from Eq.
\eqref{delta}. Since we are interested in perturbations that grow with
time, $L$ must be chosen accordingly. This can be achieved by choosing
$\frac{\eta_0}{L}\geq 1$. The value we used to obtain the solution
plotted in Fig.~\ref{fa} was $\frac{\eta_0}{L} = 10$.

According to Eq. \eqref{delta}, once we know the metric perturbation
we automatically know the time evolution of the density perturbation.
The analytical expression for this is even longer than that for the
metric perturbation so we will omit writing it down. Fig.~\ref{fb}
shows a plot of the density contrast for a selection of different
times, thus illustrating how it grows with time.

\begin{figure}
\begin{center}
\includegraphics[width=7.2cm]{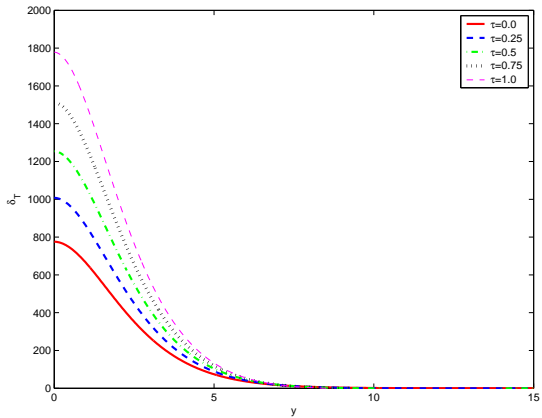} 
\end{center}
\caption{\label{fig:dT}A plot of the temperature perturbation that corresponds to
the density perturbation illustrated in Fig~\ref{fb}.}
\end{figure}

Obtaining the relative temperature change $\Delta_T$ is now simply a
matter of integrating Eq. \eqref{defofTgrad} with the density contrast
given by the analytical expression found above. Just as for the two
perturbed quantities $\delta$ and $\Phi$, we demand that the
$\Delta_T$ vanishes as $r\to\infty$. This allows us to
write the relative temperature change as
\begin{align}
	\Delta_T(y,\tau)=-&\frac{6mC}{k_BT_{00}}
	\left(\frac{L}{\eta_0}\right)^2 (1+\tau)^{\frac{2}{3}}\notag\\
	&\times
	\int_{\infty}^y\!dy'\frac{1}{y'^2}\int_0^{y'}\!dy''y''^2
	\delta(y'',\tau)\,. \label{tmppert}
\end{align}
The remaining constant which we need to determine in this expression
is the initial temperature $T_{00}$. Since the initial time and
amplitude of the density perturbation were chosen to correspond to
perturbations in our own universe at the time of recombination, it is
only natural that we also choose $T_{00}$ to be the temperature of the
universe at the same time, namely $T_{00}=3\,000\,$K. Furthermore, the
gas we consider consists of baryons, which allows us to use the mass
$m=1.67\times10^{-27}\,$kg. Using these values we find that the
dimensionless constant that multiplies the integrals in Eq.
\eqref{tmppert} will be of the order $\sim10^7$. Since the density
contrast is of the order $\sim10^{-5}$ we see that the relative change
in temperature caused by the density perturbation must be very large.
We can find a plot of this relative temperature change in
Fig~\ref{fig:dT}, which shows us that $\Delta_T$ is positive
everywhere and that it grows with time.  Thus, inserting the
analytical expression for $\Delta_T$ into Eq. \eqref{Spert}, we see
that the entropy change induced by the density perturbation will be
positive, and it will grow with time as the density inhomogeneity
increases. This shows that the entropy of a gravitationally collapsing
gas evolves according to the second law of thermodynamics up to first
order. It is natural to assume that this will be the case to any
order. 

Strictly speaking, we have only showed this for the special initial
density perturbation described in Eq. \eqref{initd}. In the appendix we
show that this will be the case for all initial density perturbations
of the class defined in the beginning of this section.
\section{Conclusions}
The main objective of this paper was to show that the entropy of an
inhomogeneous gas increases as the gas collapses under the influence
of gravity. Naively, one might expect the opposite to be the case since
inhomogeneities in both the density and the temperature increase under
such a collapse, which is an evolution that we generally associate
with a decrease in entropy. By allowing for a transfer of energy
from the gravitational potential to thermal energy, the temperature in
the gas will increase as a result of the collapse. Treating the
inhomogeneous gas as a first order perturbation to a homogeneous FRW
model, we showed that the increase in temperature results in an
increase in the entropy which outweighs any decrease due to increasing
inhomogeneities in the temperature and and the density. This was shown
to be the case for any initial density inhomogeneity which consists of
one overdense region. Although our results were derived only up to
first order in the inhomogeneity, it is only natural to extend the
conclusions to any type of inhomogeneity. This allows us to conjecture
that the entropy of a any gravitationally collapsing gas will always
increase with time, in accordance with the second law of
thermodynamics.

\begin{acknowledgments}
We wish to thank professor K.P. Tod for a very useful correpsondence
\end{acknowledgments}
\appendix*
\section{More general perturbations}
We define a class of perturbations where the initial profile has only
one overdensity. The coordinate system is chosen so that the center of
the overdensity is situated at the origin. Define the function
\begin{equation} \label{defF}
	F(y,\tau)=\int_0^{y}\!dy'y'^2\delta(y',\tau)\,.
\end{equation}
From Eq. \eqref{TotEnergyCons} we get that
\begin{equation} \label{limF}
	\lim_{y\to\infty}F(y,\tau)=0\,. 
\end{equation}
The fact that the only overdensity is that at the origin implies that
there exists a $y_0$ such that $\delta(y,\tau)>0$ for $0\leq y<y_0$
and $\delta(y,\tau)\leq0$ for $y>y_0$. This along with Eq.
\eqref{limF} implies that $F(y,\tau)$ must be positive for all values
of $y$,
\begin{equation} \label{F>0}
	F(y,\tau)>0\quad\mbox{ for } y>0\,.
\end{equation}
The temperature change in Eq. \eqref{tmppert} can be written as
\begin{equation} \label{tmppertF}
	\Delta_T(y,\tau)\propto -(1+\tau)^{2/3}\int_{\infty}^y\!dy'
	\frac{F(y',\tau)}{y'^2}\,.
\end{equation}
From this expression we see why the lower limit of the integration
must indeed be $+\infty$: We require the temperature change to
vanish at infinity. Since the integrand is always greater than zero,
this can only be accomplished if the range of integration vanishes at
infinity, which implies that the lower integration limit must be
$+\infty$. This in turn implies that the integral in Eq.
\eqref{tmppertF} will be negative for all $y$. Thus we have shown that
$\Delta_T$ will be positive for all $y$. This is the first step in our
proof. We must also show that $\Delta_T$ increases with time. In order
to do this we differentiate Eq. \eqref{tmppertF} with respect to
$\tau$. This yields the expression
\begin{align}
	\frac{\partial\Delta_T}{\partial\tau}\propto-\left(\frac{2}{3}
	(1+\tau)^{-1/3}\int_{\infty}^y\!dy'\frac{F(y',\tau)}{y'^2}\right.
	\notag\\
	+\left.(1+\tau)^{2/3}\int_{\infty}^y\!\frac{dy'}{y'^2}
	\frac{\partial F(y',\tau)}{\partial\tau}\right).
	\label{diffDelta}
\end{align}
The first term inside the parentheses will be negative due to the same
arguments as above. Using the definition in Eq. \eqref{defF}, we can
write
\begin{equation} \label{diffD2}
	H(y,\tau)\equiv\frac{\partial
	F(y,\tau)}{\partial\tau}=\int_0^{y}\!dy'y'^2
	\frac{\partial\delta(y',\tau)}{\partial\tau}\,.
\end{equation}
Again, using Eq. \eqref{limF}, we find that
$\lim_{y\to\infty}H(y,\tau)=0$. We know that the effect of gravity on
the density perturbation is such that overdense regions become more
dense, while underdense regions become less dense. This means that 
$\frac{\partial\delta(y,\tau)}{\partial\tau}>0$ for $0\leq y<y_0$
and $\frac{\partial\delta(y,\tau)}{\partial\tau}\leq0$ for $y>y_0$.
Thus, just as for the integrand in Eq. \eqref{tmppertF}, this implies
that the integrand in Eq. \eqref{diffD2} must be positive and hence
that $H(y,\tau)$ must also be positive. The integral in the second
term in  Eq. \eqref{diffDelta} must therefore be negative since the
integrand is positive while the integration path is negative. This
proves that $\frac{\partial\Delta_T}{\partial\tau}$ is positive for
all $y>0$ and $\tau>0$.

In summary, we have shown that a density perturbation of the class
defined at the start of this appendix yields a temperature change
$\Delta_T$ which is positive everywhere and grows with time. This
concludes our proof.
 

\begin{thebibliography}{11}
\expandafter\ifx\csname natexlab\endcsname\relax\def\natexlab#1{#1}\fi
\expandafter\ifx\csname bibnamefont\endcsname\relax
  \def\bibnamefont#1{#1}\fi
\expandafter\ifx\csname bibfnamefont\endcsname\relax
  \def\bibfnamefont#1{#1}\fi
\expandafter\ifx\csname citenamefont\endcsname\relax
  \def\citenamefont#1{#1}\fi
\expandafter\ifx\csname url\endcsname\relax
  \def\url#1{\texttt{#1}}\fi
\expandafter\ifx\csname urlprefix\endcsname\relax\def\urlprefix{URL }\fi
\providecommand{\bibinfo}[2]{#2}
\providecommand{\eprint}[2][]{\url{#2}}

\bibitem[{\citenamefont{Spergel et~al.}(2003)}]{spergel03}
\bibinfo{author}{\bibfnamefont{D.}~\bibnamefont{Spergel}} \bibnamefont{et~al.},
  \bibinfo{journal}{Astrophys. J.} \textbf{\bibinfo{volume}{148}},
  \bibinfo{pages}{175} (\bibinfo{year}{2003}), \eprint{astro-ph/0302209}.

\bibitem[{\citenamefont{Penrose}(1977)}]{penrose77}
\bibinfo{author}{\bibfnamefont{R.}~\bibnamefont{Penrose}}, in
  \emph{\bibinfo{booktitle}{Proceedings of the First Marcel Grossmann Meeting
  on General Relativity}}, edited by
  \bibinfo{editor}{\bibfnamefont{R.}~\bibnamefont{Ruffini}}
  (\bibinfo{publisher}{Elsevier North-Holland}, \bibinfo{year}{1977}).

\bibitem[{\citenamefont{Penrose}(1979)}]{penrose79}
\bibinfo{author}{\bibfnamefont{R.}~\bibnamefont{Penrose}}, in
  \emph{\bibinfo{booktitle}{General Relativity. An Einstein centenary survey}},
  edited by \bibinfo{editor}{\bibfnamefont{S.}~\bibnamefont{Hawking}}
  \bibnamefont{and} \bibinfo{editor}{\bibfnamefont{W.}~\bibnamefont{Israel}}
  (\bibinfo{publisher}{Cambrigde}, \bibinfo{year}{1979}).

\bibitem[{\citenamefont{Penrose}(1981)}]{penrose81}
\bibinfo{author}{\bibfnamefont{R.}~\bibnamefont{Penrose}}, in
  \emph{\bibinfo{booktitle}{Quantum gravity 2: A second Oxford symposium}},
  edited by \bibinfo{editor}{\bibfnamefont{C.}~\bibnamefont{Isham}},
  \bibinfo{editor}{\bibfnamefont{R.}~\bibnamefont{Penrose}}, \bibnamefont{and}
  \bibinfo{editor}{\bibfnamefont{D.}~\bibnamefont{Sciama}}
  (\bibinfo{publisher}{Oxford: Clarendon Press}, \bibinfo{year}{1981}).

\bibitem[{\citenamefont{Kittel and Kroemer}(1980)}]{kittel80}
\bibinfo{author}{\bibfnamefont{C.}~\bibnamefont{Kittel}} \bibnamefont{and}
  \bibinfo{author}{\bibfnamefont{H.}~\bibnamefont{Kroemer}},
  \emph{\bibinfo{title}{Thermal Physics}} (\bibinfo{publisher}{W.H. Freeman},
  \bibinfo{year}{1980}), \bibinfo{edition}{2nd} ed.

\bibitem[{\citenamefont{Mukhanov et~al.}(1992)\citenamefont{Mukhanov, Feldman,
  and Brandenberger}}]{mukhanov92}
\bibinfo{author}{\bibfnamefont{V.}~\bibnamefont{Mukhanov}},
  \bibinfo{author}{\bibfnamefont{H.}~\bibnamefont{Feldman}}, \bibnamefont{and}
  \bibinfo{author}{\bibfnamefont{R.}~\bibnamefont{Brandenberger}},
  \bibinfo{journal}{Phys.Rep.} \textbf{\bibinfo{volume}{215}}
  (\bibinfo{year}{1992}).

\bibitem[{\citenamefont{Ma and Bertschinger}(1995)}]{ma95}
\bibinfo{author}{\bibfnamefont{C.-P.} \bibnamefont{Ma}} \bibnamefont{and}
  \bibinfo{author}{\bibfnamefont{E.}~\bibnamefont{Bertschinger}},
  \bibinfo{journal}{Ap.J} \textbf{\bibinfo{volume}{455}}, \bibinfo{pages}{7}
  (\bibinfo{year}{1995}), \eprint{astro-ph/9506072}.

\bibitem[{\citenamefont{Brandenberger}(2004)}]{brandenberger03}
\bibinfo{author}{\bibfnamefont{R.}~\bibnamefont{Brandenberger}},
  \bibinfo{journal}{Lect.Notes Phys} \textbf{\bibinfo{volume}{646}},
  \bibinfo{pages}{127} (\bibinfo{year}{2004}), \eprint{hep-th/0306071}.

\bibitem[{\citenamefont{Davies}(1974)}]{davies74}
\bibinfo{author}{\bibfnamefont{P.}~\bibnamefont{Davies}},
  \emph{\bibinfo{title}{The Physics of Time Asymmetry}}
  (\bibinfo{publisher}{Surrey University Press}, \bibinfo{year}{1974}).

\bibitem[{\citenamefont{Misner et~al.}(1973)\citenamefont{Misner, Thorne, and
  Wheeler}}]{misner73}
\bibinfo{author}{\bibfnamefont{C.}~\bibnamefont{Misner}},
  \bibinfo{author}{\bibfnamefont{K.}~\bibnamefont{Thorne}}, \bibnamefont{and}
  \bibinfo{author}{\bibfnamefont{J.}~\bibnamefont{Wheeler}},
  \emph{\bibinfo{title}{Gravitation}} (\bibinfo{publisher}{Freeman},
  \bibinfo{year}{1973}).

\bibitem[{\citenamefont{Carroll and Ostlie}(1996)}]{carroll96}
\bibinfo{author}{\bibfnamefont{B.}~\bibnamefont{Carroll}} \bibnamefont{and}
  \bibinfo{author}{\bibfnamefont{D.}~\bibnamefont{Ostlie}},
  \emph{\bibinfo{title}{Modern Astrophysics}}
  (\bibinfo{publisher}{Addison-Wesley}, \bibinfo{year}{1996}).

\end{thebibliography}

\end{document}